\documentclass[amsmath,amssymb,prc,final,showpacs,twocolumn]{revtex4}
\usepackage{graphicx}  
\usepackage{bm}  
\usepackage{amsmath}
\usepackage{epsf}
\newcommand{\beq}{\begin{equation}}
\newcommand{\eeq}{\end{equation}}
\newcommand{\bea}{\begin{eqnarray}}
\newcommand{\eea}{\end{eqnarray}}

\begin{document}
\title{The Three-Nucleon System at Next-To-Next-To-Leading Order}
\author{L. Platter}\email{lplatter@phy.ohiou.edu}
\affiliation{Department of Physics and Astronomy, Ohio University,
Athens, OH 45701, USA}
\date{\today}
\begin{abstract}
We calculate higher order corrections for the three-nucleon system
up to next-to-next-to-leading within an effective field theory with
contact interactions alone.
We employ a subtraction formalism previously developed and for which it has
been shown that no new three-body force counterterm is needed for
complete renormalization up to this order. We give results for the
neutron-deuteron phaseshifts and the triton binding energy. Our results
are in very good agreement with experimental results and calculations
using realistic nucleon-nucleon potentials.
\end{abstract}


\maketitle
\section{Introduction}
The effective field theory (EFT) with contact interactions alone
allows for a systematic calculation of low-energy
two-body observables in terms of the effective range parameters \cite{Beane:2000fx}.
With the correct power counting it can be applied to systems with large
scattering length $a$ and is an expansion in $R/a$, where $R$ denotes the
range of the underlying interaction.\\
When applied to the three-body system with large scattering length,
a three-body force \cite{Bedaque:1998kg} -- or equivalently one subtraction --
is needed to obtain cutoff independent results at leading order (LO).
Although this fact seems to limit the predictive power of the EFT, it has been
very successful in describing various atomic and nuclear low-energy
observables and has been used as a tool to understand universal properties
of few-body systems with large scattering length
\cite{Bedaque:1999ve,Braaten:2004rn,Platter:2004zs}.\\ 
To achieve high precision in three-body calculations higher order corrections
have to be included. It is clear how to do this for the two-body system,
however, in the three-body system it is not a priori obvious at which order
an additional three-body counterterm has to be included.
While Hammer and Mehen \cite{Hammer:2001gh} showed that no
additional three-body datum is needed for renormalization at next-to-leading
order (NLO) and later works have come to the same conclusion, different
conclusion have been reached for different renormalization methods at
next-to-next-to-leading order (NNLO).
While Bedaque {\it et al.} \cite{Bedaque:2002yg} and
Barford and Birse \cite{Barford:2004fz} found that an additional
energy-dependent three-body counterterm is needed for renormalization
if an explicit three-body force is used and the cutoff $\Lambda$ is
kept at $\sim 1/R$ , we showed recently \cite{Platter:2006ev} that within
a subtraction formalism previously developed \cite{Hammer:2000nf,Afnan:2003bs},
no additional three-body counterterm has to be introduced at NLO and NNLO
for a consistent renormalization of observables if $\Lambda \gg 1/R$.
In this work, we computed also effective range corrections
for the $^4$He trimer system up to NNLO and achieved very good agreement with
a previous calculation using a realistic atom-atom potential. We showed that
observables are cutoff independent for sufficiently large cutoffs and that
effective range corrections to observables scale as expected.
In this paper we will extend this work to
the three-nucleon system and discuss the corresponding results. 
As the power counting in the neutron-deuteron quartet channel is well
understood \cite{Griesshammer:2005ga}, we will focus here on the
neutron-deuteron doublet channel.\\
In the following we will explain briefly the subtraction scheme which is used
in this work to renormalize three-body observables, give results for the triton
binding energy and scattering phaseshifts and discuss corrections up to NNLO
to a universal correlation between the nucleon-deuteron scattering length and
the triton binding energy, known as the Phillips line. We will end this
paper with a short conclusion and an outlook.
\section{Theory}
At sufficiently low energies few-body systems interacting through short-range
interactions can be described with an EFT built up from contact interactions
alone. Employing an auxiliary field formalism \cite{Bedaque:1998kg} the neutron-deuteron
system in the singlet channel can be described by the Lagrangian \cite{Bedaque:1999ve}
\bea
\mathcal{L}&=&N^\dagger(i\partial_0+\frac{\nabla^2}{2M})N
-t_i^\dagger(\partial_0+\frac{\nabla^2}{4M}-\Delta_t)t_i\nonumber\\
&&\qquad\qquad\qquad-s_j^\dagger(\partial_0+\frac{\nabla^2}{4M}-\Delta_s)s_j
\nonumber\\
&&\quad-G_3 N^\dagger\biggl(g_t^2(t_i\sigma_i)^\dagger t_{i'}\sigma_{i'}
\nonumber\\
&&\quad\qquad+\frac{1}{3}g_t g_s[(t_i\sigma_i)^\dagger s_j\tau_j+\text{h.c}]
\nonumber\\
&&\qquad\qquad+g_s^2(s_j\tau_j)^\dagger s_{j'}\tau_{j'}\biggr)+\ldots~,
\eea
where $N$ represents the nucleon field and $t_i (s_j)$ are the dinucleon
fields for the $^3S_1(^1S_0)$ channels with the corresponding
quantum numbers, respectively. The dots indicate more terms with more
fields/derivatives.\\
The full two-body propagator $\tau$ is the result
of dressing the bare dinucleon propagator by nucleon loops to all orders
\beq
\tau_{\alpha}(E)=-\frac{2}{\pi m}\,\frac{1}{-\gamma_\alpha+\sqrt{-mE}
+\frac{r_\alpha}{2}\left(\gamma_\alpha^2+mE\right)}~,
\label{eq:TBamp}
\eeq
where the index $\alpha=s,t$ indicates either the singlet or tiplet
two-nucleon scattering channel.
In this form the the two-body propagator has poles at energies outside the validity
region of the EFT. Therefore the propagator cannot be used within the
three-body integral equation for cutoffs $\Lambda > 1/r$ without employing
additional techniques to subtract these unphysical poles. Instead of using
the propagator in the form above we will expand it up to a given order
in $R/a$ 
\beq
\label{eq:tau}
\tau^{(n)}(E)=\frac{S^{(n)}(E)}{E+B_d},
\eeq
with $B_d=\gamma_t^2/m$ the deuteron binding energy
and for $n<3$ $S^{(n)}$ is defined as
\beq
S^{(n)}(E)=\frac{2}{\pi m^2} \sum_{i=0}^n \left(\frac{r}{2}\right)^i
[\gamma + \sqrt{-mE}]^{i+1}~.
\label{eq:Sn}
\eeq
The set of integral equations for nucleon-deuteron scattering 
generated by this EFT (neglecting the
three-body force for the beginning) is given by \cite{Bedaque:1999ve,Afnan:2003bs}
\bea
\nonumber
K_{tt}^{(n)}(q,q';E)&=&\mathcal{Z}_{tt}(q,q';E)
\\
\nonumber
&&\quad+\mathcal{P}\int_0^\Lambda\hbox{d}q''\,q''^2
\mathcal{Z}_{tt}(q,q'';E)
\\
\nonumber
&&\qquad\times\tau_t^{(n)}(E-\frac{3}{4}\frac{q''^2}{m})K_{tt}^{(n)}(q'',q';E)\\
\nonumber
&&\quad+\mathcal{P}\int_0^\Lambda\hbox{d}q''\,q''^2
\mathcal{Z}_{ts}(q,q'';E)
\\
\nonumber
&&\qquad\times\tau_s^{(n)}(E-\frac{3}{4}\frac{q''^2}{m})K_{st}^{(n)}(q'',q';E)~,
\eea
\bea
\label{eq:integraleq}
\nonumber
K_{st}^{(n)}(q,q';E)&=&\mathcal{Z}_{st}(q,q';E)
\\
\nonumber
&&\quad+\mathcal{P}\int_0^\Lambda\hbox{d}q''\,q''^2
\mathcal{Z}_{st}(q,q'';E)
\\
\nonumber
&&\qquad\times\tau_t^{(n)}(E-\frac{3}{4}\frac{q''^2}{m})K_{tt}^{(n)}(q'',q';E)\\
\nonumber
&&\quad+\mathcal{P}\int_0^\Lambda\hbox{d}q''\,q''^2
\mathcal{Z}_{ss}(q,q'';E)
\\
&&\qquad\times\tau_s^{(n)}(E-\frac{3}{4}\frac{q''^2}{m})K_{st}^{(n)}(q'',q';E)~,
\nonumber\\
\eea
where $m$ denotes the nucleon, $n$ the order of the calculation (for
the following definitions it will always be assumed that $n<3$)
and  $\mathcal{Z}_{\alpha\beta}$ the Born amplitude
\beq
\mathcal{Z}_{\alpha\beta}(q,q';E)=-\lambda_{\alpha\beta}\frac{m}{q q'}
\log\bigl(\frac{q^2+q q'+q'^2-mE}{q^2-q q'+q'^2-mE}\bigr)~,
\eeq
with the iso-spin matrix $\lambda$ given by
\beq
\lambda=\frac{1}{4}\left(\begin{array}{cc}
1 & -3  \\
-3 & 1  \end{array}\right)~ 
\eeq
The set of integral equations in Eq.(\ref{eq:integraleq}) is strongly cutoff
dependent and a three-body force has to be introduced or equivalently
a subtraction has to be performed to render observables cutoff independent.
At threshold the integral equations are renormalized by noting that
\beq
K_{tt}^{(n)}(0,0;-B_d)=\frac{3 m a_3}{8\gamma\sum_{i=0}^n(\gamma r)^n}~,
\eeq
and subtracting this known quantity from Eq.(\ref{eq:integraleq})
\bea
\nonumber
K_{tt}^{(n)}(q,0;-B_d)&-&K_{tt}^{(n)}(0,0;-B_d)~,\\
K_{st}^{(n)}(q,0;-B_d)&-&K_{tt}^{(n)}(0,0;-B_d)~.
\eea
After rewriting the resulting set of integral equations the half-off-shell
threshold amplitude takes the following form \cite{Hammer:2000nf,Afnan:2003bs}
\bea
K_{tt}^{(n)}(q,0;-B_d)&=&K_{tt}^{(n)}(0,0;-B_d)
\nonumber\\
&&+[\mathcal{Z}_{tt}(q,0;-B_d)-\mathcal{Z}_{tt}(0,0;-B_d)]
\nonumber
\\
&&+\int_0^\Lambda\hbox{d}q''\,q''^2
[\mathcal{Z}_{tt}(q,q'';-B_d)
\nonumber\\
&&\quad\qquad\qquad-\mathcal{Z}_{tt}(0,q'';-B_d)]
\nonumber
\\
&&\times\tau_t^{(n)}(-B_d-{\textstyle\frac{3q''^2}{4m}})K_{tt}^{(n)}(q'',0;-B_d)
\nonumber
\\
&&+\int_0^\Lambda\hbox{d}q''\,q''^2
[\mathcal{Z}_{ts}(q,q'';-B_d)
\nonumber\\
&&\quad\qquad\qquad-\mathcal{Z}_{ts}(0,q'';-B_d)
\nonumber
\\
&&\times\tau_s^{(n)}(-B_d-{\textstyle\frac{3q''^2}{4m}})K_{st}^{(n)}(q'',0;-B_d)~,
\nonumber
\eea
\bea
K_{st}^{(n)}(q,0;-B_d)&=&K_{tt}^{(n)}(0,0;-B_d)
\nonumber\\
&&+[\mathcal{Z}_{st}(q,0;-B_d)-\mathcal{Z}_{tt}(0,0;-B_d)]
\nonumber
\\
&&+\int_0^\Lambda\hbox{d}q''\,q''^2
[\mathcal{Z}_{st}(q,q'';-B_d)
\nonumber\\
&&\quad\qquad\qquad-\mathcal{Z}_{tt}(0,q'';-B_d)]
\nonumber
\\
&&\times\tau_t^{(n)}(-B_d-{\textstyle\frac{3q''^2}{4m}})K_{tt}^{(n)}(q'',0;-B_d)
\nonumber
\\
&&+\int_0^\Lambda\hbox{d}q''\,q''^2
[\mathcal{Z}_{ss}(q,q'';-B_d)
\nonumber\\
&&\qquad\quad\qquad-\mathcal{Z}_{ts}(0,q'';-B_d)]
\nonumber
\\
&&\times\tau_s^{(n)}(-B_d-{\textstyle\frac{3q''^2}{4m}})K_{st}^{(n)}(q'',0;-B_d)~.
\nonumber\\
\eea
The amplitudes $K_{tt}^{(n)}(q,0;-B_d)$ and $K_{st}^{(n)}(q,0;-B_d)$ are
fully renormalized after the subtraction is performed. An essential point
in obtaining the amplitudes at any energy and momentum is demanding that
\bea
K_{tt}^{(n)}(q,0;-B_d)&=&K_{tt}^{(n)}(0,q;-B_d)~,
\nonumber
\\
K_{ts}^{(n)}(q,0;-B_d)&=&K_{st}^{(n)}(0,q;-B_d)~.
\eea
Using resolvent identities, subtracted integral equations for any
energy can be derived. For further details, the reader is advised to
turn to \cite{Afnan:2003bs,Platter:2006ev}.\\
The two-body parameters used throughout the rest of this work are given
by
\bea
\gamma_t^{-1}&=&4.317~\hbox{fm}~,\qquad r_t=1.764~\hbox{fm}~,
\nonumber
\\
\gamma_s^{-1}&=&-25.04~\hbox{fm}~,\qquad r_s=2.73~\hbox{fm}~.
\eea
We will use the result of a recent neutron-deuteron scattering length
measurement with $a_3=0.645 \pm 0.005$~fm as our three-body
input \cite{Huffman:2005jx}.
\section{Results}
\begin{figure}[b]
\centerline{\includegraphics*[width=8cm,angle=0]{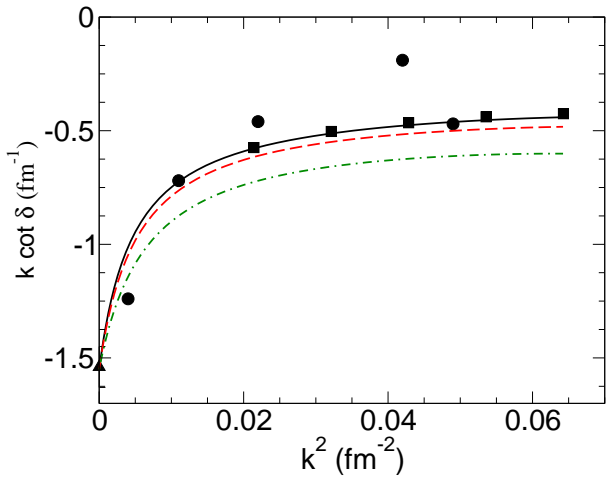}}
\caption{\label{fig:nd_kcotd}(Color online) Phase shifts for neutron-deuteron scattering
below the deuteron breakup at LO (dashed-dotted line), NLO (dashed line)
and NNLO (solid line). The triangle is the result of the scattering
length measurement of \cite{Huffman:2005jx}. The circles are the results
of the van Oers-Seagrave phaseshift analysis \cite{seagrave}, and the squares denote
a phaseshift calculation using a realistic nucleon-nucleon potential~\cite{Kievsky:1996ca}}
\end{figure}
The three-nucleon system has been previously considered within the EFT
with contact interactions alone. Leading order results for phaseshift
and the correlation between triton binding energy and neutron-deuteron
scattering length were obtained in
\cite{Bedaque:1999ve}. Bedaque {\it et al.} \cite{Bedaque:2002yg}
calculated higher order corrections up to NNLO introducing
an additional energy-dependent three-body force at this order. In this
section we will demonstrate by explicit calculation that no additional
three-body force is needed for renormalization and considerable
improvement in results is achieved at NNLO. 
We have computed the triton binding energy at LO, NLO and NNLO using the
neutron-deuteron scattering length as the three-body input parameter
\cite{Huffman:2005jx}. Our results
are displayed in table~\ref{table:tritonenergy}. When using the experimental
value for the neutron-deuteron singlet scattering length $a_3=0.645$~fm as 
three-body input we obtain a triton binding energy of $B_3=8.48$~MeV at NNLO
which has to be compared to the experimental value $B_3^{\text{Exp}}=8.54$~MeV.
\begin{table}[t]
\begin{center}
\begin{tabular}{|c|c||c|c||}
\hline
&parameters& $a_3$ [fm] & $B_3$ [Mev]\\
\hline \hline
 LO  &$a_3$,$\gamma_t$,$a_s$& 0.645      & 8.08\\
NLO  &$a_3$,$\gamma_t$,$a_s$,$r_t$,$r_t$& 0.645      & 8.19\\
NNLO &$a_3$,$\gamma_t$,$a_s$,$r_t$,$r_t$& 0.645      & 8.54\\      
\hline
EXP & &0.65      & 8.48\\
\hline
\end{tabular}
\end{center}
\caption{\label{table:tritonenergy} EFT predictions for the triton
binding energy up to NNLO using the neutron-deuteron scattering
length as three-body input. The second column indicates which parameters were
used at what order of the calculation. Energies and lengths are given in MeV and
fm, respectively.
}
\end{table}
The results show a significant improvement with each order and are
sufficiently close at NNLO to the experimental value to agree with a
projected error at this order of approximately $(r_t \gamma_t)^3\sim$ 3\%.
At these orders the error caused by the uncertainty in the effective
range parameters is smaller than the error caused by N$^3$LO corrections.\\
Using the set of integral equations in Eq.(\ref{eq:integraleq}) with
the expanded two-body propagator in Eq.(\ref{eq:tau}) leads at LO, NLO
and NNLO to the phaseshifts shown in Fig.\ref{fig:nd_kcotd}.
For comparison we show in the same figure the results of a forty year old
phaseshift analysis \cite {seagrave} and a theoretical calculation using a realistic
nucleon-nucleon potential \cite{Kievsky:1996ca}. At higher order our
results seem to describe the experimental data better but considering
the age of the analysis and the fact that no errors are given for these data,
the theoretical calculation  by Kievsky {\it et al.} should be
considered as the true benchmark test for our calculation.   
At NLO our results already lie significantly closer to this calculation
and nearly perfect agreement is achieved at NNLO.
It should be noted that our results at LO and NLO order agree
with previous EFT calculations results given in
\cite{Hammer:2001gh,Bedaque:2002yg,Afnan:2003bs}. We also achieve qualitative
agreement at NNLO with Ref.\cite{Bedaque:2002yg}, however, without
employing an additional three-body counterterm.

\begin{figure}[t]
\centerline{\includegraphics*[width=8cm,angle=0]{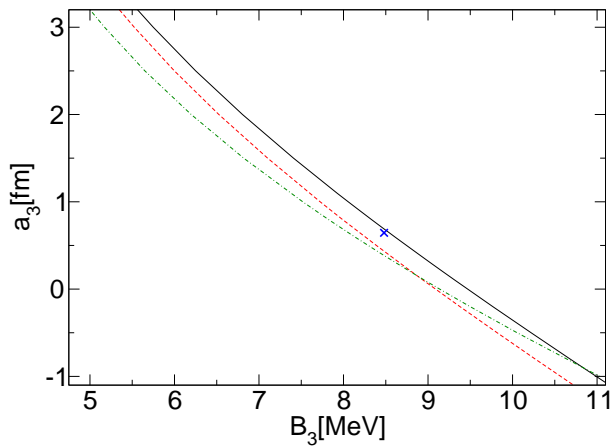}}
\caption{\label{fig:phillipsline} (Color online) The Phillips line for leading (dot-dashed
  line), next-to-leading (dashed line) and next-to-next-to-leading
(solid line) order. The cross denotes the experimental value.}
\end{figure}

A further way to illustrate the improvement in our results is to consider
the Phillips line. The Phillips line is a universal feature of three-body
systems with a large two-body scattering length and arises as a nearly
linear correlation between the 1+2 scattering length and the three-body
binding energy. In Fig.\ref{fig:phillipsline} we display our results
for LO, NLO and NNLO and display also the experimental value.
It is interesting to note
that in contradistinction to the $^4$He trimer system the Phillips line
has not converged to a definite result yet \cite{Platter:2006ev}. The obvious reason for this is the
rather large expansion coefficient which is roughly $\gamma_t r_t \sim 1/3$
while $\gamma r \sim 0.1$ in the $^4$He system.
The power counting derived in \cite{Platter:2006ev} is valid for cutoffs
much larger than $1/r$. Therefore, our results have been evaluated at
corresponding cutoffs and in fact our results are fully converged
with respect to $\Lambda$. The results given in this section are
also numerically converged up to the digits displayed.
Further, we also analyzed the convergence of three-body
observables when using a different parametrisation for
higher order two-body corrections called Z-matching
\cite{Phillips:1999hh,Griesshammer:2004pe} and have found
the same results up to the level of accuracy expected in an NNLO
calculation.
\section{Summary}
In this paper we have computed three-nucleon observables with the
EFT with contact interactions alone up to NNLO using a subtraction scheme
for renormalization.   We have shown that the results for observables
improve significantly at NNLO without performing an additional subtraction,
{\it i,e.} without the introduction of a further energy dependent three-body
counterterm.
This improvement is in particular obvious if one considers the results for the
binding energies and scattering phaseshifts simultaneously. Our value
for the triton binding energy agrees very well with the experimental value
and the results for the neutron-deuteron phaseshifts seem to be in nearly perfect
agreement with a calculation using a realistic nucleon-nucleon potential.
The results for the scattering phaseshifts agree qualitatively
with previous calculations at NLO \cite{Hammer:2001gh,Afnan:2003bs} and
NNLO \cite{Bedaque:2002yg}, although a second three-body counterterm
was included in \cite{Bedaque:2002yg}.
In particular, these results seem to agree very well with an expected error
of $(\gamma r)^3\sim 3\%$ for an NNLO calculation. We have therefore presented
further numerical evidence which supports the claim in Ref.~\cite{Platter:2006ev}
that calculations can be performed with exactly on three-body counterterm
up to NNLO in the EFT with contact interactions alone.
\\
Our analysis presented in \cite{Platter:2006ev} indicates that at N$^3$LO
an additional three-body input is needed for renormalization of observables.
Therefore, NNLO can be also considered as the last order at
which a prediction can be made for the Phillips line as it is a
correlation gouverned by one three-body parameter.

Further possible applications of the subtraction formalism include the
calculation of scattering observables above the breakup threshold and the
coupling of external currents to the three-nucleon system, including
the electromagnetic form factor beyond leading order \cite{Platter:2005sj}.

\begin{acknowledgments}
I thank H.-W.~Hammer and D.~R.~Phillips
for stimulating discussions and comments on the manuscript.
Furthermore, I thank Chubu University
for its hospitality during partial completion of this work.
This work was supported by the U.S. Department of Energy under
grant DE-FG02-93ER40756.
\end{acknowledgments}
\bibliographystyle{prsty}

\end{document}